\documentclass[sigconf]{acmart}

\AtBeginDocument{%
  }

\copyrightyear{2025}
\acmYear{2025}
\setcopyright{cc}
\setcctype{by}
\acmConference[MM '25]{Proceedings of the 33rd ACM International Conference on Multimedia}{October 27--31, 2025}{Dublin, Ireland}
\acmBooktitle{Proceedings of the 33rd ACM International Conference on Multimedia (MM '25), October 27--31, 2025, Dublin, Ireland}
\acmDOI{10.1145/3746027.3754868}
\acmISBN{979-8-4007-2035-2/2025/10}

\usepackage{algorithmic}
\usepackage{graphicx}
\usepackage{textcomp}
\usepackage{xcolor}
\usepackage{makecell}
\usepackage{subcaption}
\usepackage{float}
\usepackage{booktabs}
\usepackage{multirow}
\usepackage{amsmath}
\usepackage[skip=3pt]{caption}
\usepackage{enumitem} 
\usepackage{balance}

\begin{document}

\title{REMOTE: A Unified Multimodal Relation Extraction Framework with Multilevel Optimal Transport and Mixture-of-Experts}

\author{Xinkui Lin}
\email{linxinkui@iie.ac.cn}
\affiliation{
  \institution{Institute of Information Engineering, Chinese Academy of Sciences}
  \institution{School of Cyber Security, University of Chinese Academy of Sciences}
  \city{Beijing}
  \country{China}
}

\author{Yongxiu Xu}
\authornote{Yongxiu Xu and Hongbo Xu are joint corresponding authors.}
\email{xuyongxiu@iie.ac.cn}
\affiliation{
  \institution{Institute of Information Engineering, Chinese Academy of Sciences}
  \institution{School of Cyber Security, University of Chinese Academy of Sciences}
  \city{Beijing}
  \country{China}
}

\author{Minghao Tang}
\email{tangminghao@iie.ac.cn}
\affiliation{
  \institution{Institute of Information Engineering, Chinese Academy of Sciences}
  \institution{School of Cyber Security, University of Chinese Academy of Sciences}
  \city{Beijing}
  \country{China}
}

\author{Shilong Zhang}
\email{zhangshilong@iie.ac.cn}
\affiliation{
  \institution{Institute of Information Engineering, Chinese Academy of Sciences}
  \institution{School of Cyber Security, University of Chinese Academy of Sciences}
  \city{Beijing}
  \country{China}
}

\author{Hongbo Xu}
\authornotemark[1] 
\email{hbxu@iie.ac.cn}
\affiliation{
  \institution{Institute of Information Engineering, Chinese Academy of Sciences}
  \institution{School of Cyber Security, University of Chinese Academy of Sciences}
  \city{Beijing}
  \country{China}
}

\author{Hao Xu}
\email{xuhao@iie.ac.cn}
\affiliation{
  \institution{Institute of Information Engineering, Chinese Academy of Sciences}
  \institution{School of Cyber Security, University of Chinese Academy of Sciences}
  \city{Beijing}
  \country{China}
}

\author{Yubin Wang}
\email{wangyubin@iie.ac.cn}
\affiliation{
  \institution{Institute of Information Engineering, Chinese Academy of Sciences}
  \institution{School of Cyber Security, University of Chinese Academy of Sciences}
  \city{Beijing}
  \country{China}
}


\renewcommand{\shortauthors}{Xinkui Lin et al.}

\begin{abstract}
Multimodal relation extraction (MRE) is a crucial task in the fields of Knowledge Graph and Multimedia, playing a pivotal role in multimodal knowledge graph construction.
However, existing methods are typically limited to extracting a single type of relational triplet, which restricts their ability to extract triplets beyond the specified types. 
Directly combining these methods fails to capture dynamic cross-modal interactions and introduces significant computational redundancy.
Therefore, we propose a novel \textit{unified multimodal Relation Extraction framework with Multilevel Optimal Transport and mixture-of-Experts}, termed REMOTE, which can simultaneously extract intra-modal and inter-modal relations between textual entities and visual objects. 
To dynamically select optimal interaction features for different types of relational triplets, we introduce mixture-of-experts mechanism, ensuring the most relevant modality information is utilized. 
Additionally, considering that the inherent property of multilayer sequential encoding in existing encoders often leads to the loss of low-level information, we adopt a multilevel optimal transport fusion module to preserve low-level features while maintaining multilayer encoding, yielding more expressive representations. 
Correspondingly, we also create a Unified Multimodal Relation Extraction (UMRE) dataset to evaluate the effectiveness of our framework, encompassing diverse cases where the head and tail entities can originate from either text or image.
Extensive experiments show that REMOTE effectively extracts various types of relational triplets and achieves state-of-the-art performanc on almost all metrics across two other public MRE datasets.
We release our resources at \url{https://github.com/Nikol-coder/REMOTE}.
\end{abstract}




\maketitle

\section{Introduction}
Multimodal relation extraction (MRE)~\cite{9428274, 10.1145/3474085.3476968, 10.1145/3477495.3531992, 10.1145/3589334.3645603, 10.1145/3664647.3681219, hu-etal-2023-multimodal,10.1145/3581783.3612209,10.1145/3664647.3681367,10.1145/3627673.3679883} aims to leverage text and image information to extract relational triplets, represented as (\textit{head entity, tail entity, relation}).
With the rapid growth of multimedia content online, MRE has gained increasing significance, driving advancements in cross-modal retrieval and multimodal knowledge graph construction.

Existing MRE methods can be broadly divided into two main categories based on the types of relational triplets they extract:
Multimodal Neural Relation Extraction (MNRE)~\cite{9428274, 10.1145/3474085.3476968, 10.1145/3477495.3531992, 10.1145/3589334.3645603, 10.1145/3664647.3681219, hu-etal-2023-multimodal,ZHAO2023103264,li-etal-2023-dual-gated}, which leverages images as auxiliary information to extract relations between textual entities, and Multimodal Object-Entity Relation Extraction (MORE)~\cite{10.1145/3581783.3612209}, which incorporates images into relational triples to extract relations between textual entities and visual objects.

\begin{figure}[t]
\centering
    \Description{Examples of different multimodal relation extraction tasks.} 
  \includegraphics[width=0.49\textwidth]{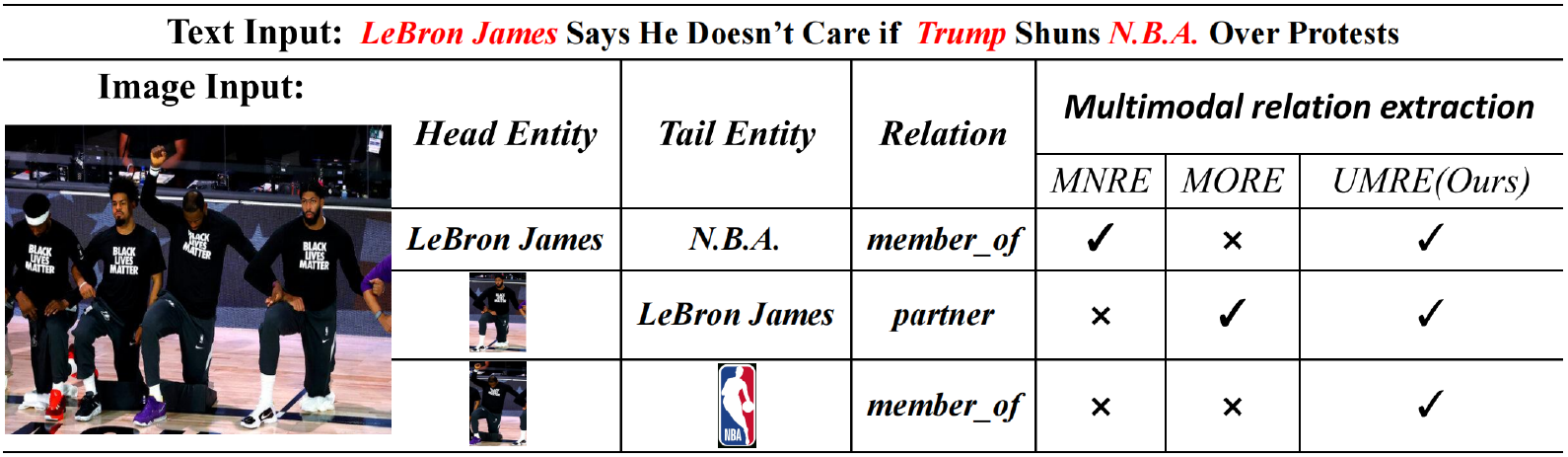}
    \caption{Examples of different multimodal relation extraction tasks. Symbols $\checkmark$ and $\times$ indicate whether the task successfully extracts the relational triplet or not.}
  \label{fig:task}
\end{figure}

Although these methods have spurred significant advancements, they are typically constrained to extracting a single type of relational triplet, limiting their ability to extract triplets beyond the specified types.
As shown in Fig.~\ref{fig:task}, MNRE methods fail to extract relations involving visual objects, while MORE methods are restricted to extracting relations solely between textual entities and visual objects.  
This task-specific design, coupled with the absence of dynamic feature selection mechanisms for different types of relational triplets, severely limits their effectiveness in real-world applications requiring adaptive multimodal reasoning.

To address this limitation, we propose a unified multimodal relation extraction framework, capable of simultaneously extracting intra-modal (between textual entities, between visual objects) and inter-modal (between textual entities and visual objects) relations.
Specifically, we employ Mixture-of-Experts (MoE) mechanism to dynamically select the optimal interaction features for different types of relational triplets. By leveraging visual, textual, and multimodal experts, the MoE mechanism fully utilizes the most relevant information from each modality.

Furthermore, the inherent property of multilayer sequential encoding in existing encoders often leads to the loss of low-level information~\cite{cao2024mmfusermultimodalmultilayerfeature,10.5555/3666122.3667638,pmlr-v234-yu24a}, which hinders the comprehensive representations of modality information.
Therefore, we propose a multilevel optimal transport fusion module to preserve low-level features, ensuring both high-level semantic information and low-level details are effectively captured for relation extraction.

To evaluate our method, we also create the first unified multimodal relation extraction dataset called \textbf{UMRE}. 
It covers three types of multimodal relational triplets, extracting intra-modal and inter-modal relations between textual entities and visual objects, surpassing previous MRE datasets in both scale and diversity.
Extensive experiments on the UMRE and two other public MRE benchmark datasets demonstrate the effectiveness of our method.

To sum up, our contributions are summarized as follows:
\begin{itemize}[topsep=0pt, itemsep=0pt, parsep=0pt, partopsep=0pt]
    \item To the best of our knowledge, we are the first to propose Unified Multimodal Relation Extraction, which simultaneously extracts intra-modal and inter-modal relations between textual entities and visual objects. In addition, we construct a high-quality dataset for this task, surpassing existing datasets in both scale and diversity.
    \item We propose a unified multimodal relation extraction framework that employs Mixture-of-Experts mechanism to dynamically select the most relevant interaction features for different types of relational triplets. 
    Moreover, the framework leverages optimal transport to fuse multilevel feature, capturing both high-level semantic information and low-level details.
    \item Extensive experiments on UMRE and two widely-used MRE datasets demonstrate the superiority of our method.
\end{itemize}

\section{Related Works}
\subsection{Multimodal Relation Extraction}
Multimodal relation extraction, which aims to leverage image and text information to extract relational triplets, has attracted significant attention in recent years.
Early works treated images as auxiliary information, primarily extracting relations between textual entities. 
BERT+SG~\cite{10.1145/3474085.3476968} utilized the entire image to enhance semantic information and extracted relational triplets between textual entities.
However, image and text are not always fully related, and using the entire image without filtering may introduce substantial noise irrelevant to the text.
To address this issue, MEGA~\cite{9428274} leveraged scene graphs to align texts and images, focusing on image regions relevant to text to reduce noise.

Recent methods like MKGformer~\cite{10.1145/3477495.3531992} and HVFormer~\cite{10.1145/3589334.3645603} utilized dual-modality alignment to leverage hierarchical visual context, further eliminating noise.
FocalMRE~\cite{10.1145/3664647.3680995} introduced focal attention mechanisms to emphasize text-related image regions.
As multimedia develops, relational triplets are evolving from unimodal to multimodal.
MOREformer~\cite{10.1145/3581783.3612209} introduced a new task called multimodal object-entity relation extraction (MORE), which incorporated images into relational triplets to extract relations between textual entities and visual objects.

Despite these advancements, existing methods are designed for extracting a single type of relational triplet, restricting their ability to handle diverse triplets beyond the specified types.
Hence, we propose a unified multimodal relation extraction framework that dynamically selects the most relevant features for different types of relational triplets, enabling extracting simultaneously intra-modal and inter-modal relations between textual entities and visual objects.

\section{Task Definition}

Building on the MNRE~\cite{10.1145/3474085.3476968} and MORE~\cite{10.1145/3581783.3612209} tasks, we integrate various Multimodal Relation Extraction tasks into a unified paradigm, referred to as the \textbf{Unified Multimodal Relation Extraction (UMRE)} task. 
As illustrated in Fig.~\ref{fig:task}, UMRE extends traditional multimodal relation extraction by simultaneously capturing three distinct relational triplets based on both textual entity set \(e_{set}\) and visual object set \(o_{set}\) from text \(T\) and image \(I\): (1) intra-modal relations between textual entities, (2) intra-modal relations between visual objects, and (3) inter-modal relations between textual entities and visual objects. 
The task can be formally defined as follows:
\begin{equation}
(T, I, e_{set}, o_{set}) \rightarrow R 
\end{equation}

Where the relation set \( R \) include:
\begin{equation}
R = \{(\mathit{e}, \mathit{e}, \mathit{r}) \cup (\mathit{o}, \mathit{o}, \mathit{r}) \cup (\mathit{e}, \mathit{o}, \mathit{r})\} 
\end{equation}

Here, \((\mathit{e}, \mathit{e}, \mathit{r})\) denotes relations between textual entities, \((\mathit{o}, \mathit{o}, \mathit{r})\) denotes relations between visual objects, and \((\mathit{e}, \mathit{o}, \mathit{r})\) denotes relations between textual entities and visual objects, where textual entities \( \mathit{e} \in e_{set} \) and visual objects \( \mathit{o} \in o_{set} \).

\section{Dataset Construction}

\begin{figure}[htb]
  \Description{The distribution of relations.} 
  \includegraphics[width=0.78\columnwidth]{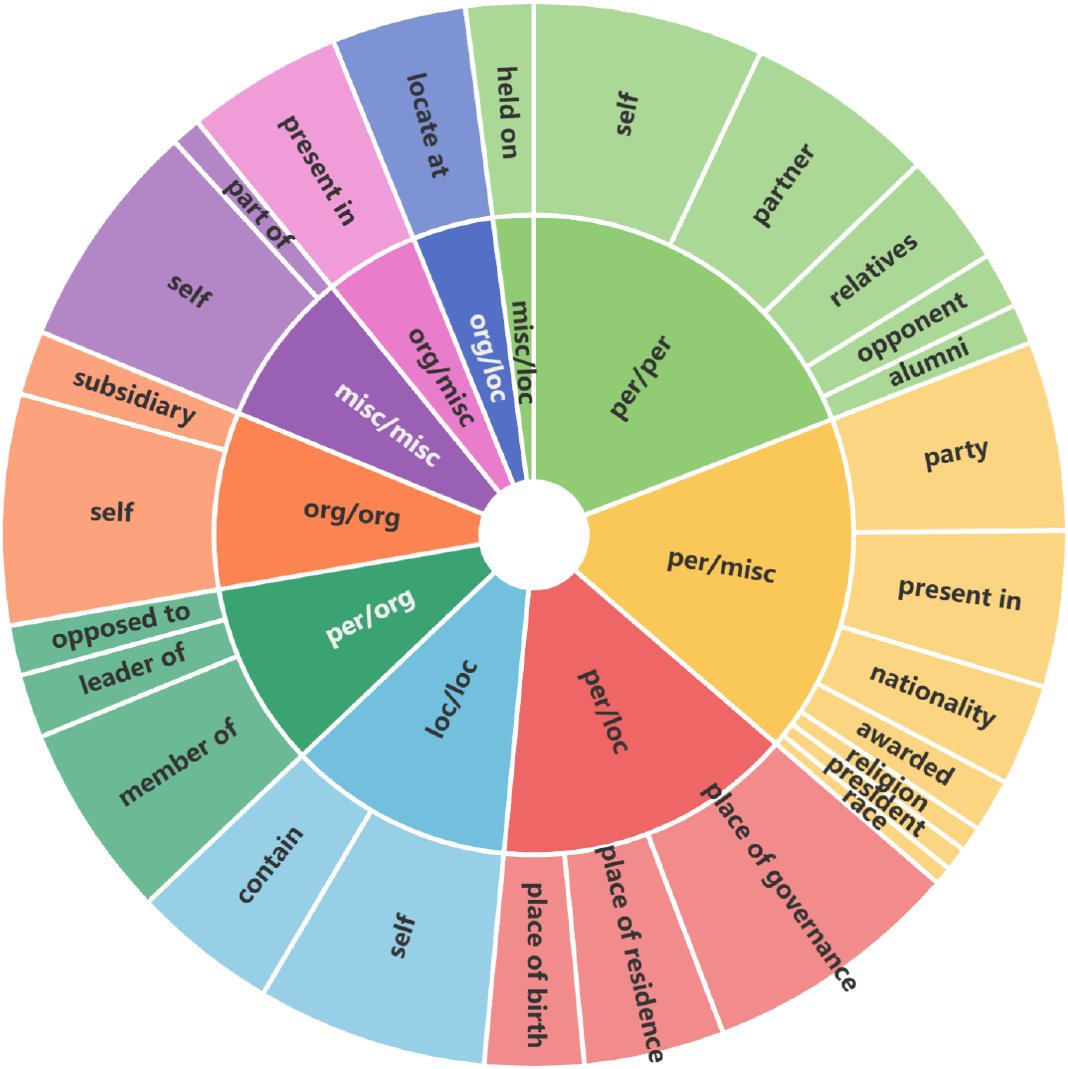}
  \caption{The Distribution of Relation Categories in Our UMRE Dataset.}
  \label{fig:umke_relationships}
\end{figure}

\begin{table}[htb]
\centering
\caption{Comparison of UMRE with existing RE datasets.\newline (MM: Multimodal, Img: Images, Sent: Sentences, VO: Visual Objects, Fact: Number of relational triplets, Type: Number of relational triplet Types, Rel: Number of Relations)}
\label{table:dataset_comparison}
\resizebox{\columnwidth}{!}{
\begin{tabular}{cccccccc}
\hline
\textbf{Dataset} & \textbf{MM} & \textbf{Img} & \textbf{Sent} & \textbf{VO} & \textbf{Fact} & \textbf{Type} & \textbf{Rel} \\ 
\hline
ACE 03-04\cite{doddington-etal-2004-automatic}  & No & - & 12,783 & - & 16,771  & 1 & 24  \\
TACRED\cite{alt-etal-2020-tacred}     & No & - & 53,791 & - & 21,773 & 1 & 41   \\
FewRel\cite{gao-etal-2019-fewrel}     & No & - & 56,109 & - & 70,000 & 1 & 100  \\
MNRE\cite{10.1145/3474085.3476968}      & Yes & 9,201 & 9,201 & - & 15,485 & 1 & 23  \\
MORE\cite{10.1145/3581783.3612209}       & Yes & 3,559 & 3,559 & 13,520 & 20,264 & 1 & 21  \\ 
\textbf{UMRE(Ours)}  & \textbf{Yes} & \textbf{12,737} & \textbf{12,737} & \textbf{20,978} & \textbf{55,021} & \textbf{3} & \textbf{28}  \\
\hline
\end{tabular}
}
\end{table}

To evaluate the effectiveness of our framework, we introduce a new dataset named UMRE for unified multimodal relation extraction.
Table~\ref{table:dataset_comparison} presents a comprehensive comparison of UMRE with several existing datasets for relation extraction, including ACE 2003-2004~\cite{doddington-etal-2004-automatic}, TACRED~\cite{alt-etal-2020-tacred}, FewRel~\cite{gao-etal-2019-fewrel}, multimodal relation extraction dataset MNRE~\cite{10.1145/3474085.3476968} and MORE~\cite{10.1145/3581783.3612209}.

The UMRE dataset builds upon the MORE and MNRE datasets, aiming to extract intra-modal and inter-modal relations between textual entities and visual objects. 
UMRE encompasses 28 relation types (shown in Fig.~\ref{fig:umke_relationships}), with 55,021 annotated multimodal relational triplets derived from 12,737 text-image pairs, including 3,890 triplets between visual objects, 15,438 triplets between textual entities, and 35,693 triplets between textual entities and visual objects. 
Additionally, we introduce new self-relations to reflect cases where textual entities and visual objects refer to the same entities. 

To ensure high-quality multimodal relation annotations, the UMRE dataset construction follows a rigorous two-stage pipeline that synergizes automated extraction with human verification. 

\subsection{Stage 1: Named Entity Recognition and Object Detection}
We extend the MORE and MNRE datasets by leveraging Multimodal Large Language Models (MLLMs), such as Qwen2-VL~\cite{wang2024qwen2vlenhancingvisionlanguagemodels}, Qwen2.5-VL~\cite{bai2025qwen25vltechnicalreport} and LLAMA3.2-Vision~\cite{grattafiori2024llama3herdmodels}, to identify and extract potential candidate entities and objects from both text and image data. 

Following RIVEG~\cite{li-etal-2023-prompting}, we first construct a Support Set by selecting 200 representative samples from the MNRE and MORE training sets to enhance the recognition process, ensuring coverage of diverse text entities and visual objects. For each sample in the UMRE dataset, we calculate the multimodal representation similarity between the sample and the Support Set using CLIP~\cite{DBLP:conf/icml/RadfordKHRGASAM21}. Then, we select the top 5 most relevant samples from the Support Set as few-shot examples and feed them into the MLLM. 

The MLLM extracts potential candidate entities from the text and provides relevant information about these entities, which is integrated with the original text. This mixed information is processed by a fine-tuned RoBERTa model \cite{li-etal-2023-prompting, li-etal-2024-llms, li2024advancinggroundedmultimodalnamed} to determine the definitive entities. Additionally, the MLLM describes the spatial positions of these entities within images, and LLAVA \cite{liu2024llavanext} processes these descriptions to obtain precise bounding box coordinates. Finally, all annotated objects and entities are manually reviewed and corrected by our annotators. 

\subsection{Stage 2: Multimodal Relation Annotation}
We recruit six educated annotators to analyze image and text information, identifying intra-modal and inter-modal relations between textual entities and visual objects. Triplets without explicit relations are labeled as none. 
All relational triplets are independently reviewed by at least two annotators. Because ambiguous hierarchical relations exist, there may be discrepancies in annotation results. An independent adjudicator was introduced to finalized annotations. 
The Weighted Cohen's Kappa evaluates the consistency between different annotators\footnote{The Kappa value is 0.7325 here.}. 

\section{Our Proposed Method}

\begin{figure*}[htb]
   \Description{The framework of REMOTE.} 
   \centering
  \includegraphics[width=\textwidth]{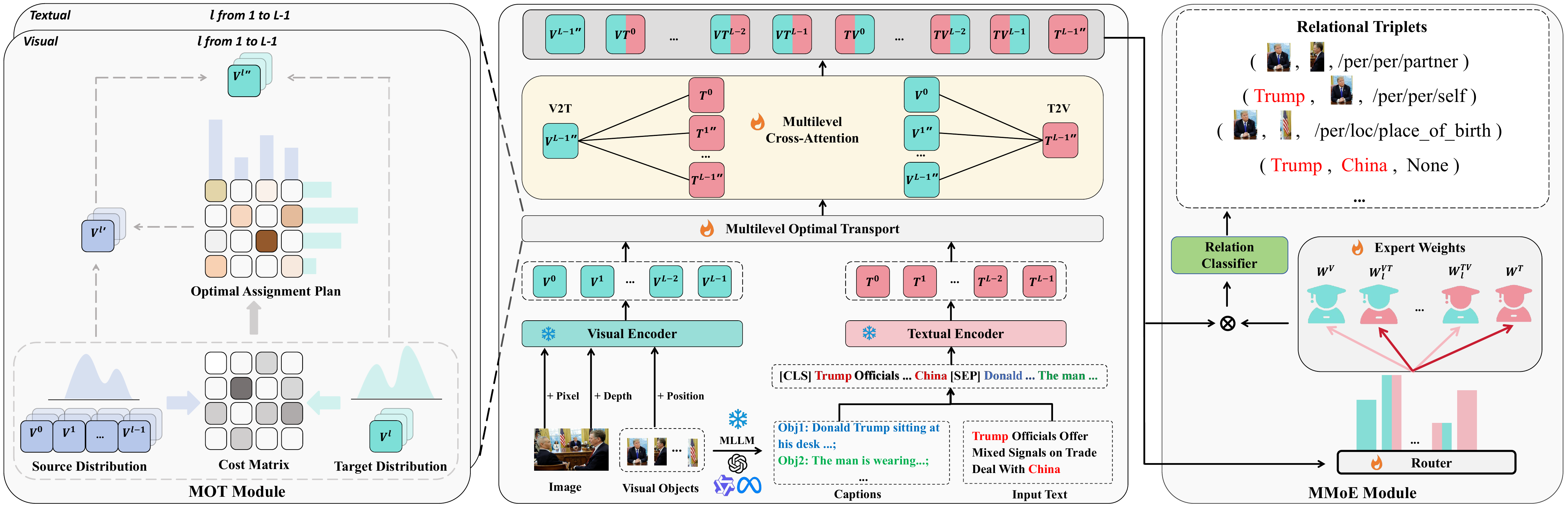}
  \caption{The framework of REMOTE. Texts with different colors use different separators.}
  \label{fig:model}
\end{figure*}

In this section, we present a unified multimodal relation extraction framework, named REMOTE, as shown in in Fig.~\ref{fig:model}.  
The framework is designed to effectively extracts intra-modal and inter-modal relations between textual entities and visual objects.
It comprises three key components: (\ref{sec:feature_encoding}) Feature Encoding, (\ref{sec:multilevel_optimal_transport}) Multilevel Optimal Transport, and (\ref{Multimodal_Mixture_of_Experts}) Multimodal Mixture-of-Experts.
Firstly, the Feature Encoding module processes both visual and textual inputs separately. The visual encoder extracts features from images and visual objects, while the textual encoder processes captions and input text to generate their respective representations.  
Secondly, the Multilevel Optimal Transport module aligns these multilevel features by constructing an optimal assignment plan, producing more comprehensive representations.  
Finally, the Multimodal Mixture-of-Experts module dynamically selects optimal interaction features for different types of relational triplets, ensuring the most relevant modality information is utilized. 
Each component plays a crucial role in effectively capturing and integrating multimodal information for multimodal relation extraction tasks.

\subsection{Feature Encoding}
\label{sec:feature_encoding}
\subsubsection{Textual Encoding}

To bridge the semantic gap between text and image, we leverage multimodal large language models\footnote{To enhance the robustness of our approach, we use multiple MLLMs, including BLIP2, Qwen2-VL, Qwen2.5-VL, and Llama3.2-Vision.} (MLLM) to generate captions $T_{i}^{cap} = \mathrm{MLLM}(o_i)$ for visual objects $o_i$, where $i \in [0, 12]$. These captions are segmented using the special token \(\langle o \rangle\), while textual entities \(x_m\) in the original text are segmented with \(\langle s \rangle\). Next, the full text sequence is encapsulated with $[CLS]$ and $[SEP]$ tokens as input to BERT~\cite{devlin-etal-2019-bert}, producing the textual features \(F_T \in \mathbb{R}^{n \times d_t}\), where \(n\) denotes the sequence length and \(d_t\) is the dimension of hidden states of BERT. The process is as follows:

\begin{equation}
\begin{aligned}
{F}_{T}^{cap} &= \{ \langle o \rangle T_{i}^{cap} \langle /o \rangle \} \\
{F}_{T}^{org} &= \{ x_1, \dots, \langle s \rangle x_m^1, \dots, x_m^{{len}(x_m)} \langle /s \rangle, \dots, x_n \} \\
{F}_{T} &= \{ [CLS], F_{\!T}^{org}, [SEP], F_{\!T}^{cap} \}
\end{aligned}
\end{equation}

\subsubsection{Visual Encoding}

Following MOREformer~\cite{10.1145/3581783.3612209}, visual objects that are closer to the center of the image, occupy larger proportions, and are positioned closer to the foreground are more likely to represent primary objects and exhibit stronger associations with text content.
We utilize the Vision Transformer (ViT)~\cite{dosovitskiy2021imageworth16x16words} and the Depth-Anything model~\cite{yang2024depthv2} to generate more comprehensive visual features by integrating RGB, depth, and position information.
Visual objects are segmented by their spatial coordinates, while Depth-Anything encodes the depth information. 
We scale the RGB value \(V_{RGB}\) and depth value \(V_{DEPTH}\) of each visual object \(o_i\) to unified \(H \times W\) pixels, and feed them into ViT for encoding, producing patch embeddings \(E_{RGB} \in \mathbb{R}^{u \times d_v}\)  and  \(E_{DEPTH} \in \mathbb{R}^{u \times d_v}\), where \( u = \frac{H \times W}{P^2} \), \( P \) is the patch size, and \( d_v \) represents the dimension of hidden states of ViT. Finally, we concatenate the position embeddings \(E_{POS} \in \mathbb{R}^{u \times d_v}\) with \(E_{RGB}\) and \(E_{DEPTH}\) to form the visual features \(F_V \in \mathbb{R}^{2u \times d_v}\). The process is as follows:

\begin{equation}
\begin{aligned}
F_V^{rgb} &= E_{RGB} \oplus E_{POS} \\
F_V^{depth} &= E_{DEPTH} \oplus E_{POS} \\
F_V &= \text{Concat}[F_V^{rgb}; F_V^{depth}] 
\end{aligned}
\end{equation}

Where \(\oplus\) denotes element-wise addition.  

\subsection{Multilevel Optimal Transport}
\label{sec:multilevel_optimal_transport}

\begin{figure}[ht]
   \Description{Visualization of attention across different layers.} %
  \begin{minipage}[b]{0.24\linewidth}
    \centering
    \includegraphics[width=\linewidth,height=0.07\textheight]{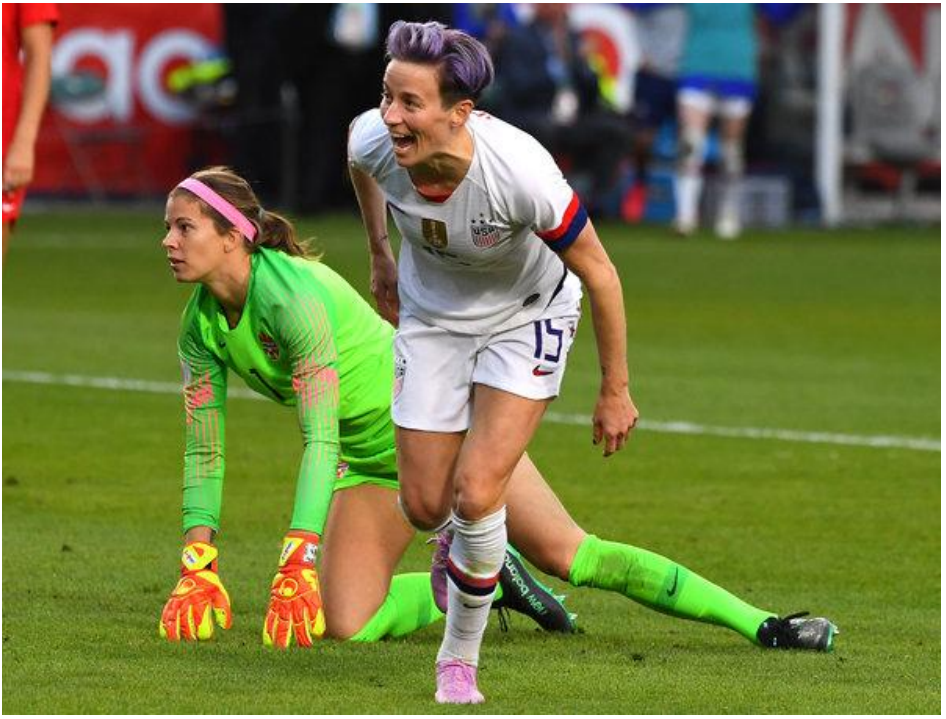}
    \subcaption*{Original}
  \end{minipage}
  \hfill
  \begin{minipage}[b]{0.24\linewidth}
    \centering
    \includegraphics[width=\linewidth,height=0.07\textheight]{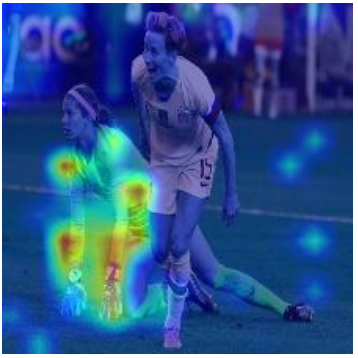}
    \subcaption*{ViT(Layer 4)}
  \end{minipage}
  \hfill
  \begin{minipage}[b]{0.24\linewidth}
    \centering
    \includegraphics[width=\linewidth,height=0.07\textheight]{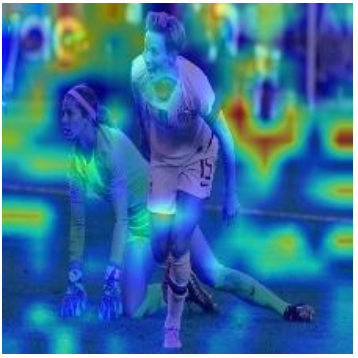}
    \subcaption*{ViT(Layer 7)}
  \end{minipage}
  \hfill
  \begin{minipage}[b]{0.24\linewidth}
    \centering
    \includegraphics[width=\linewidth,height=0.07\textheight]{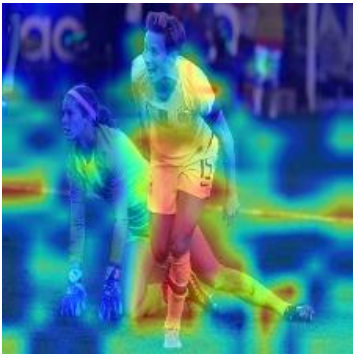}
    \subcaption*{ViT(Layer 12)}
  \end{minipage}
  \caption{Visualization of attention across different layers. (red: high attention, blue: low attention).}
  \label{fig:attention_visualization}
\end{figure}

Previous methods~\cite{10.1145/3474085.3476968,10.1145/3477495.3531992,10.1145/3589334.3645603,10.1145/3664647.3681219,hu-etal-2023-multimodal,10.1145/3581783.3612209} usually utilize visual features from the final or penultimate layer of the ViT and BERT, effectively leveraging high-level semantic features to align the image and text modalities.
However, the inherent property of multilayer sequential encoding in existing encoders often leads to the loss of low-level information~\cite{10.5555/3666122.3667638,pmlr-v234-yu24a}, which may hinder the correct extraction of relational triplets that rely on these details.
Specifically, higher layers focus on the main objects of the image, while lower layers capture finer details, such as scene and object attributes~\cite{cao2024mmfusermultimodalmultilayerfeature}. As illustrated in Fig.~\ref{fig:attention_visualization}, Layer 12 highlights the primary object in the image, Layer 7 captures broader scene information like the football field, and Layer 4 emphasizes finer details such as the goalkeeper.

To address this issue, we propose a Multilevel Optimal Transport fusion module (MOT, as shown in Fig.~\ref{fig:model}). This module leverages Optimal Transport (OT) theory~\cite{moradi2025surveyalgorithmicdevelopmentsoptimal,10476763} to quantify differences by minimizing transport costs between different distributions. 
By utilizing Sinkhorn-optimized alignment~\cite{10.5555/2999792.2999868}, our method efficiently integrates hierarchical representations from various encoder layers. This integration results in more comprehensive representations while maintaining computational efficiency. 

For visual features extracted from ViT, we denote the first \( l \) layers as \(F_{V^{low}} = (F_{V^{0}}, \dots, F_{V^{l-1}})\), where each layer's representation \(F_{V^l} \in \mathbb{R}^{2u \times d_v}\) corresponds to spatial semantic features at depth \( l \) (\( 0 < l \leq L-1 \)), and \( L \) represents the total number of layers in the encoder.  
These lower-level features are concatenated to form the source distribution \(\mu \in \mathbb{R}^{(l \cdot 2u) \times d_v}\), capturing fine-grained details. Meanwhile, the higher-level feature \(F_{V^{l}} \in \mathbb{R}^{2u \times d_v}\) from layer \( l \) serves as the target distribution \(\nu \in \mathbb{R}^{2u \times d_v}\), which encodes global semantics but lacks local details.  

The distributions are formally defined as:  
\begin{equation}  
\begin{cases}
\text{Source Distribution: } \mu = \text{Concat}[F_{V^0}, \dots, F_{V^{l-1}}] \\
\text{Target Distribution: } \nu = F_{V^l}
\end{cases}
\end{equation}  

Then, we compute the minimal total transportation cost \(\mathbf{\Pi}^\star\) between \({\mu}\) and \({\nu}\), which aligns cross-level features while preserving their marginal distributions. To improve computational efficiency, we introduce the entropy regularization~\cite{10.5555/2999792.2999868} to smooth the optimization problem:

\begin{equation}
\begin{aligned}
& \mathbf{\Pi}^\star = \arg\min_{\mathbf{\Pi} \in \mathbb{R}_+^{a \times b}} \sum_{i=1}^a \sum_{j=1}^b C_{ij} \mathbf{\Pi}_{ij} - \lambda H(\mathbf{\Pi}) \\
& \mathbf{s.t.} \quad  \mathbf{\Pi} \mathbf{1}_b = {\mu}, \quad \mathbf{\Pi}^\top \mathbf{1}_a = {\nu} \\
\end{aligned}
\end{equation}

Where \( \mathbf{1}_b \in \mathbb{R}^b \) and \( \mathbf{1}_a \in \mathbb{R}^a \) are all-ones vectors, and \( \lambda \) controls the entropy regularization strength. Respectively, \( b = l \cdot 2u \) and \( a = 2u \) denote the number of feature elements in \({\mu}\) and \({\nu}\), ensuring dimensional consistency in transportation planning.

\begin{equation}
\begin{aligned}
C_{ij} &= 1 - \frac{{\mu}_i^\top {\nu}_j}{\|{\mu}_i\| \cdot \|{\nu}_j\|} \\
H(\mathbf{\Pi}) &= -\sum_{i,j} \mathbf{\Pi}_{ij} \log(\mathbf{\Pi}_{ij}) \\
\end{aligned}
\end{equation}

Here, \(C \in \mathbb{R}^{m \times n}\) denotes the cosine similarity cost between \( {\mu}_{i} \) and \( {\nu}_{j} \), \( H(\mathbf{\Pi}) \) is the entropy term. The optimal transport plan \( \mathbf{\Pi}^\star \) is efficiently solved via the Sinkhorn-Knopp algorithm~\cite{10.5555/2999792.2999868}.

Finally, the transported features \( F_{V^{l'}} \) are fused with the original \( F_{V^{l}} \) via a learnable weight \( {\alpha} \in \mathbb{R}^{d_v} \), yielding the enhanced representation \( F_{V^{l''}} \).  

\begin{equation}
\begin{aligned}
   F_{V^{l'}} &= \mathbf{\Pi}^\star \cdot {\mu} \\
    F_{V^{l''}} &= {\alpha} \odot F_{V^{l}} + (1 - {\alpha}) \odot F_{V^{l'}}
\end{aligned}
\end{equation}  

Where \( \odot \) denotes element-wise multiplication. This adaptive fusion balances semantic richness and detail preservation.  

Similarly, textual features are processed through the same procedure to produce an enhanced representation \( F_{T^{l''}} \).
The MOT module not only preserves valuable fine-grained information, but also aggregates contextual semantics, effectively bridging the semantic gap between low-level details and high-level semantics.

\subsection{Multimodal Mixture of Experts}
\label{Multimodal_Mixture_of_Experts}

We propose a hierarchical cross-modal interaction method that establishes comprehensive multimodal interaction features through layer-wise attention mechanisms. Specifically, this method implements bidirectional information flow by employing high-level semantic representations from one modality as queries to hierarchically attend to multilevel features from the complementary modality. This process generates two groups of complementary hierarchical cross-modal features: \( F_{j}^{V \to T} \) (vision-guided textual features) and \( F_{j}^{T \to V} \) (text-guided visual features) at each hierarchical level \( j \in [0, L-1] \).

\begin{equation}
\scalebox{0.95}{$
\begin{gathered}
Q^{V \to T} = F_{V^{L-1''}} W_q, \quad K_j^{V \to T} = F_{T^{j''}} W_k, \quad H_j^{V \to T} = F_{T^{j''}} W_h \\
A_j^{V \to T} = \text{Softmax} ( \frac{Q^{V \to T} (K_j^{V \to T})^\top}{\sqrt{d}} ) \\
F_{j}^{V \to T} = A_j^{V \to T} H_j^{V \to T}
\end{gathered}
$}
\end{equation}

\begin{equation}
\scalebox{0.95}{$
\begin{gathered}
Q^{T \to V} = F_{T^{L-1''}} W_q, \quad K_j^{T \to V} = F_{V^{j''}} W_k, \quad H_j^{T \to V} = F_{V^{j''}} W_h \\
A_j^{T \to V} = \text{Softmax} ( \frac{Q^{T \to V} (K_j^{T \to V})^\top}{\sqrt{d}} ) \\
F_{j}^{T \to V} = A_j^{T \to V} H_j^{T \to V}
\end{gathered}
$}
\end{equation}

Where \(W_q, W_k, W_h \in \mathbb{R}\) denote learnable projection matrices.  

This hierarchical architecture facilitates progressive cross-modal information fusion, where high-level semantic features from the source modality guide attention over multilevel features from the target modality, empowering the model to bridge semantic gaps between visual-textual relationships through progressive alignment at different levels.

However, for effective multimodal relation extraction, simply aggregating these multimodal features may introduce noise and redundancy rather than benefits.
To address this, we introduce Mixture of Experts (MoE) mechanism to dynamically select the most relevant experts for each triplet, which ensures that the most critical information from each modality is effectively utilized, while reducing interference from irrelevant or noisy features.

Moreover, to mitigate the potential noise from irrelevant textual and visual information, we add textual-only and visual-only features to the cross-modal interaction features. The Multimodal Mixture of Experts module (MMoE) dynamically assigns weights to high-level textual features, high-level visual features, and cross-modal interaction features based on different relational triplets. These weights are collectively determined by visual, textual, and multimodal experts.

\begin{equation}
S(T, V) = \text{Softmax} ( [ F_{V^{L-1''}}; F_0^{V \to T}; \dots; F_{L-1}^{T \to V}; F_{T^{L-1''}} ] \cdot P )
\end{equation}

Where \( S(T, V) \) represents the expert routing mechanism, which assigns weights to each \( F(T, V)_j \) using a softmax function, with \( P \) as a trainable mapping matrix.  

The MMoE module leverages a dynamic routing mechanism to optimize the fusion of modalities. This mechanism is particularly effective in capturing complex cross-modal interactions, because it allows the model to focus on modality-specific and modality-shared representations simultaneously.

\begin{equation}
\begin{aligned}
H &= \text{MMoE}(T, V) = \sum_{j=0}^{L-1} S(T, V)_j \cdot F(T, V)_j \\
&= \sum_{j=0}^{L-1} S_j^{V \to T} \cdot F_j^{V \to T} + \sum_{j=0}^{L-1} S_j^{T \to V} \cdot F_j^{T \to V} \\
&\quad + S^{T} \cdot F_{T^{L-1''}} + S^{V} \cdot F_{V^{L-1''}} 
\end{aligned}
\end{equation}

This mechanism ensures that the model can dynamically prioritize modalities based on different relational triplets, thereby enhancing the robustness and accuracy of cross-modal reasoning.

For unified multimodal relation extraction, we utilize different interaction features based on the type of relational triplet. For textual entities, we adopt the textual features segmented by $\langle s \rangle$ after cross-model interaction as the final textual representations. For visual objects, in addition to using the textual features of corresponding captions segmented by $\langle o \rangle$ after cross-model interaction, we also incorporate the visual features from object regions to form comprehensive visual representations. 

\begin{equation} 
\begin{aligned} 
{h} &= 
\begin{cases} 
H_{\langle s \rangle}^{T,V} & \text{if } {h} \text{ is a textual entity} \\
[ H_{\langle o_i \rangle}^{T,V}, H_{o_i}^{{V},{T}} ] & \text{if } {h} \text{ is a visual object}  
\end{cases} 
\\
P_r &= \text{argmax} ( \text{MLP} ( [ {h}_{head}, {h}_{tail} ] ) ) \\
\end{aligned} 
\end{equation}

The model parameters are optimized by minimizing the cross-entropy loss between the prediction and the ground truth:

\begin{equation}
L_{umre} = -\sum_{r} y_r \log(P_r)
\end{equation}

\section{Experiments}

\begin{table*}[ht]
\caption{The overall performance of REMOTE and other state-of-the-art methods on datasets. The best results are marked in \textbf{bold}, and the second-best results are \underline{underlined}. * represents results from reproduced experiments.}
\label{table:results}
\centering
\resizebox{\textwidth}{!}{
    \begin{tabular}{c|c|cccc|cccc|cccc}
    \hline
    \multirow{2}{*}{\textbf{Model}} & \multirow{2}{*}{\textbf{Venue}} 
    & \multicolumn{4}{c|}{\textbf{UMRE}} & \multicolumn{4}{c|}{\textbf{MORE}} & \multicolumn{4}{c}{\textbf{MNRE}} \\
    \cline{3-14}
    & & Acc & Pre & Rec & F1 & Acc & Pre & Rec & F1 & Acc & Pre & Rec & F1 \\
    \hline
    BERT+SG \cite{10.1145/3474085.3476968} & ICME21 & 63.30 & 44.92 & 42.11 & 43.47 & 61.79 & 29.61 & 41.27 & 34.48 & 74.09 & 62.95 & 62.65 & 62.80  \\
    BERT+SG+Att \cite{10.1145/3474085.3476968} & ICME21 & 64.58 & 47.21 & 43.13 & 45.07 & 63.74 & 31.10 & 39.28 & 34.71 & 74.59 & 60.97 & 66.56 & 63.64  \\
    MEGA \cite{9428274} & MM21 & 65.85 & 49.51 & 47.24 & 48.34 & 65.97 & 33.30 & 38.53 & 35.72 & 76.15 & 64.51 & 68.44 & 66.41 \\
    VisualBERT \cite{li2019visualbertsimpleperformantbaseline} & arXiv19 & 58.42 & 54.07 & 53.29 & 53.68 & {82.84} & 58.18 & 61.22 & 59.66 & - & 57.15 & 59.45 & 58.30 \\
    IFAformer \cite{10.1609/aaai.v37i13.26987} & AAAI23 & 74.41 & 60.21 & 62.04 & 61.11 & 79.28 & 55.13 & 54.24 & 54.68 & 92.38 & 82.59 & 80.78 & 81.67 \\
    MKGformer \cite{10.1145/3477495.3531992} & SIGIR22 & \underline{75.77} & 60.22 & 60.26 & 60.23 & 80.17 & 55.76 & 53.74 & 54.73 & 83.36 & 82.40 & 81.73 & 82.06 \\
    MRE \cite{Hu2023mre} & ACL23 & - & - & - & - & - & - & - & - & 93.54 & 85.03 & 84.25 & 84.64 \\
    MRE-ISE \cite{wu-etal-2023-information} & ACL23 & - & - & - & - & - & - & - & - & 94.06 & 84.69 & 83.38 & 84.03 \\  
    DGF-PT \cite{10.1145/3581783.3611899} & MM23 & 74.54 & 61.54 & 61.36 & 61.44 & 82.35 & 60.51 & 62.82 & 61.64 & 84.25 & 84.35 & 83.83 & 84.47 \\
    MOREformer \cite{10.1145/3581783.3612209} & MM23 & 75.47 & \underline{63.03} & {62.77} & {62.89}  & \underline{83.50} & \underline{62.18} & \underline{63.34} & \underline{62.75} & 82.67 & 82.19 & 82.35 & 82.27\\ 
    MMIB \cite{10.1109/TASLP.2023.3345146} & TASLP24 & 74.62 & 61.58 & 60.74 & 61.16 & 82.41 & 60.15 & 62.28 & 61.17 & - & 83.49 & 82.97 & 83.23 \\
    HVFormer \cite{10.1145/3589334.3645603} & WWW24 & \underline{75.77} & 61.37 & 61.04 & 61.20 & 82.31 & 58.81 & 62.84 & 60.76 & - & {84.14} & {82.65} & {83.39} \\
    CGI-MRE \cite{wu-etal-2023-information} & ICMR24 & 74.08 & 60.45 & 61.84 & 61.63 & 81.92 & 57.44 & 63.01 & 60.09 & - & 85.02 & 84.22 & 84.62 \\ 
    GBIT \cite{10448507} & ICASSP24 & - & - & - & - & - & - & - & - & 93.55 & 85.21 & 83.75 & 84.48 \\
    FocalMRE \cite{10.1145/3664647.3680995} & MM24 & 73.82 & 61.77 & \underline{66.11} & \underline{63.87} & 82.44 & 60.75 & 62.89 & 61.81 & -/\underline{94.24*} & \textbf{88.85}/86.96* & \underline{87.19}/85.47* & \textbf{88.01}/86.21* \\ 
    \hline
    Qwen2-VL-7B \cite{wang2024qwen2vlenhancingvisionlanguagemodels} & arXiv24 & 12.23 & 7.32 & 11.52 & 8.95  & 10.44 & 8.52 & 12.87 & 10.25 & 14.52 & 11.42 & 13.09 & 12.20 \\
    Qwen2.5-VL-7B \cite{bai2025qwen25vltechnicalreport} & arXiv25 & 16.96 & 12.33 & 12.04 & 12.19  & 23.94 & 25.81 & 23.95 & 24.84 & 18.25 & 13.82 & 15.63 & 14.68 \\
    Llama-3.2-11B-Vision \cite{grattafiori2024llama3herdmodels} & arXiv24 & 9.09 & 5.98 & 13.46 & 8.28 & 6.69 & 5.09 & 18.66 & 8.01 & 11.94 & 9.89 & 13.32 & 11.35 \\
    \hline
    \textbf{Ours} & - & \textbf{78.47} & \textbf{66.92} & \textbf{71.56} & \textbf{69.17} & \textbf{83.64} & \textbf{63.21} & \textbf{64.63} & \textbf{63.91} & \textbf{94.86} & \underline{88.05} & \textbf{87.50} & \underline{87.77} \\
    $\Delta_{\text{SOTA}}$ & - & $\uparrow\mathbf{2.7}$ & $\uparrow\mathbf{3.89}$ & $\uparrow\mathbf{5.45}$ & $\uparrow\mathbf{5.3}$ & $\uparrow\mathbf{0.14}$ & $\uparrow\mathbf{1.03}$ & $\uparrow\mathbf{1.29}$ & $\uparrow\mathbf{1.16}$ & $\uparrow\mathbf{0.62}$ & $\downarrow0.8/\uparrow\mathbf{1.09}$ & $\uparrow\mathbf{0.31}/\uparrow\mathbf{2.03}$ & $\downarrow0.24/\uparrow\mathbf{1.56}$ \\
    \hline
    \end{tabular}
}
\end{table*}

\subsection{Implementation Details} 
All experiments are conducted using BERT-{base} and ViT-B/32 on a single NVIDIA RTX 4090 GPU. The dimension of textual features is set to 768, and the dimension of visual object features is set to 4096. 
The maximum numbers of token sequences and objects are limited to 128 and 12, respectively. Our model is optimized using the AdamW optimizer~\cite{loshchilov2019decoupledweightdecayregularization} with a base learning rate of 1e-5 and a batch size of 32. A dropout rate of 0.5 is applied in our experiments. We conduct 3 averages for each experiment. 

\subsection{Dataset} In addition to UMRE, we also evaluate our method on two public MRE datasets: (1) \textbf{MNRE}\cite{10.1145/3474085.3476968}, with 15,485 samples across 23 relations, focusing on extracting relations between textual entities; (2) \textbf{MORE}\cite{10.1145/3581783.3612209}, comprising 21 relations and 20,264 multimodal relational triplets, designed for extracting relations between textual entities and visual objects.

\subsection{Baseline} 
We compare our REMOTE method with state-of-the-art approaches across two categories:

\noindent\textbf{Multimodal Relation Extraction Models}:
\begin{itemize}
\item \textbf{BERT+SG+Att}~\cite{10.1145/3474085.3476968}: Concatenates textual embeddings with scene graph-derived visual features for relation extraction.  
\item \textbf{MEGA}~\cite{9428274}: Aligns textual-visual graphs through efficient cross-modal alignment strategies.  
\item \textbf{VisualBERT}~\cite{li2019visualbertsimpleperformantbaseline}: Leverages self-attention to implicitly align text tokens and image regions.  
\item \textbf{IFAformer}~\cite{10.1609/aaai.v37i13.26987}: Achieves fine-grained alignment via cross-modal transformer layers.  
\item \textbf{MKGformer}~\cite{10.1145/3477495.3531992}: Integrates modalities via hybrid Transformer architecture.  
\item \textbf{MRE}~\cite{Hu2023mre}: Injects knowledge-aware information via multimodal retrieval.  
\item \textbf{MRE-ISE}~\cite{wu-etal-2023-information}: Introduces information subtraction/addition mechanisms for noise reduction.  
\item \textbf{DGF-PT}~\cite{10.1145/3581783.3611899}: Pretrains fusion modules using self-supervised signals from image-caption pairs.  
\item \textbf{MOREformer}~\cite{10.1145/3581783.3612209}: Incorporates spatial and depth information for visual object disambiguation.  
\item \textbf{MMIB}~\cite{10.1109/TASLP.2023.3345146}: Applies information bottleneck principle to filter modality noise.  
\item \textbf{HVFormer}~\cite{10.1145/3589334.3645603}: Dynamically selects cross-modal features via Visual Mixture-of-Experts.  
\item \textbf{CGI-MRE}~\cite{wu-etal-2023-information}: Leverages genetic algorithms for cross-modal common information discovery.  
\item \textbf{GBIT}~\cite{10448507}: Combines graph neural networks and transformers for cross-modal fusion.  
\item \textbf{FocalMRE}~\cite{10.1145/3664647.3680995}: Enhances image region through focal augmentation strategies.  
\end{itemize}  

\noindent\textbf{Multimodal Large Language Models}:
\begin{itemize}  
\item \textbf{Qwen2-VL-7B}~\cite{wang2024qwen2vlenhancingvisionlanguagemodels}: Alibaba's single-stream pre-trained model, supporting native dynamic resolution.  
\item \textbf{Qwen2.5-VL-7B}~\cite{bai2025qwen25vltechnicalreport}: Enhanced with context length and improved visual reasoning for long-video understanding and object localization.  
\item \textbf{Llama3.2-11B-Vision}~\cite{grattafiori2024llama3herdmodels}: Meta's vision-augmented model for cross-modal understanding.  
\end{itemize}  

TMR~\cite{zheng-etal-2023-rethinking} and CAMIM~\cite{10.1145/3664647.3681219} employ text-to-image generation~\cite{9878449} for data augmentation. However, this method changes the original layout and positional information of image elements during generation, which makes it unsuitable for unified relation extraction tasks requiring strict spatial information preservation. Therefore, these methods are excluded from our comparative experiments.

Since the baselines were originally designed for specific multimodal relation extraction (MRE) tasks, we adapt them to evaluate performance across all datasets. For evaluation, we report accuracy, precision, recall, and F1 score. Among these metrics, we prioritize the \textbf{F1 score} as it addresses the long-tail effect and provides a robust assessment of performance on imbalanced tasks.

\subsection{Overall Results}
We conduct experiments on the three datasets, and Table \ref{table:results} illustrates the overall results. The following observations are made: 
\begin{enumerate}
    [leftmargin=*, itemsep=2pt, parsep=0pt, topsep=5pt]
    \item Multimodal large language models without fine-tuning perform poorly on MRE tasks, indicating their inadequacy in addressing MRE challenges effectively.
    \item Models like BERT+SG+Att, MEGA, and VisualBERT, which incorporate the entire image as auxiliary information, do not achieve significant improvements. 
    This indicates that the image is not always fully related to the text and often contains substantial noise, which can hinder MRE performance.
    \item Models such as MKGformer, HVFormer, MOREformer and FocalMRE, which utilize object detection and visual grounding to selectively leverage text-relevant image regions, perform notably better. This highlights the effectiveness of fine-grained image-text alignment in reducing noise from irrelevant visual information.
    \item 
    Our proposed REMOTE model outperforms all baseline models on almost all metrics across the three MRE datasets, demonstrating its superior performance. Specifically, compared to the current SOTA model, our model achieves a relative improvement of \textbf{5.3\% in F1 score} and \textbf{2.7\% in Accuracy} on the UMRE dataset. Moreover, our model shows notable performance gains on the MORE and MNRE datasets as well. These results further demonstrate that REMOTE can dynamically select the optimal interaction features for different relational triplets.
\end{enumerate}

\subsection{Ablation Study}
\begin{table}[htb]
\caption{
    The ablation study results on datasets. "w/o" indicates the removal of the corresponding module.
}
\label{table:ablation-results}
\centering
\small 
\resizebox{\columnwidth}{!}{%
    \begin{tabular}{c|cc|cc|cc}
    \hline
    \multirow{2}{*}{\textbf{Model}} & \multicolumn{2}{c|}{\textbf{UMRE}} & \multicolumn{2}{c|}{\textbf{MORE}} & \multicolumn{2}{c}{\textbf{MNRE}} \\
    \cline{2-7}
    & \textbf{Acc} &  \textbf{F1} & \textbf{Acc} & \textbf{F1} & \textbf{Acc} & \textbf{F1} \\
    \hline
    \text{MOT (Optimal Transport)}  & \textbf{78.47} & \textbf{69.17} & \textbf{83.64} & \textbf{63.91} & \textbf{94.86} & \textbf{87.77} \\
    \text{MOT (Cross Attention)}  & 78.01 & 68.23 & 82.31 & 63.03  & 94.36 & 86.72 \\ 
    \hline
    \text{w/o MOT}  & 75.64 & 64.82 & 82.80 & 61.68  & 93.68 & 85.08 \\ 
    \hline
    \text{w/o MMoE}  & 76.61 & 66.48 & 82.52 & 62.01  & 94.23 & 86.21 \\
    \hline
    \end{tabular}
    }
\end{table}

\begin{figure*}[ht]
  \centering
  \Description{Multimodal expert weights visualization results.}
  
  \subfloat[Misc/Loc/held on]{
    \includegraphics[width=0.3\linewidth, height=4.5cm, keepaspectratio]
    {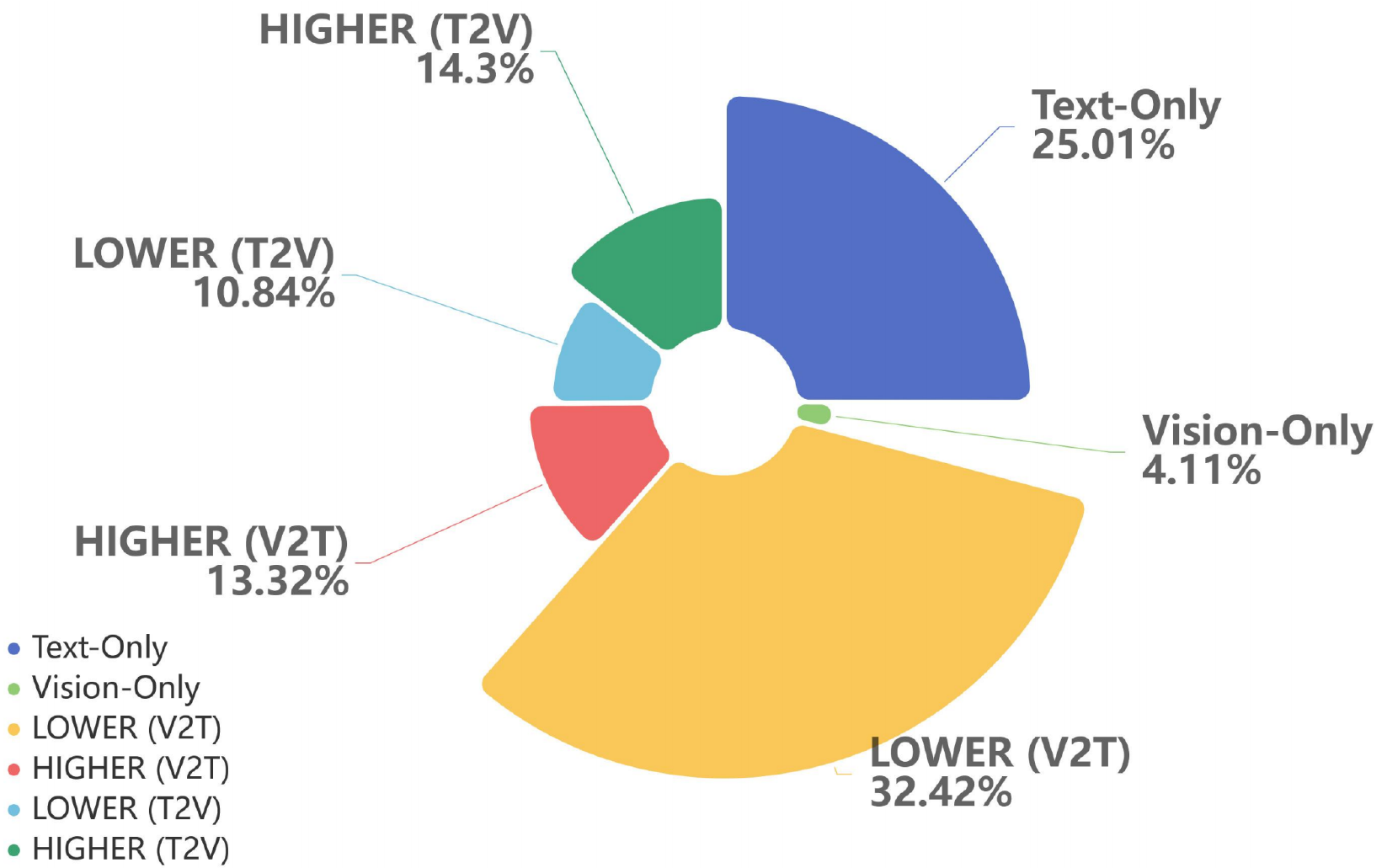}
    \label{subfig:case1}
  }\hfill
  \subfloat[Per/Misc/president]{
    \includegraphics[width=0.3\linewidth, height=4.5cm, keepaspectratio]
    {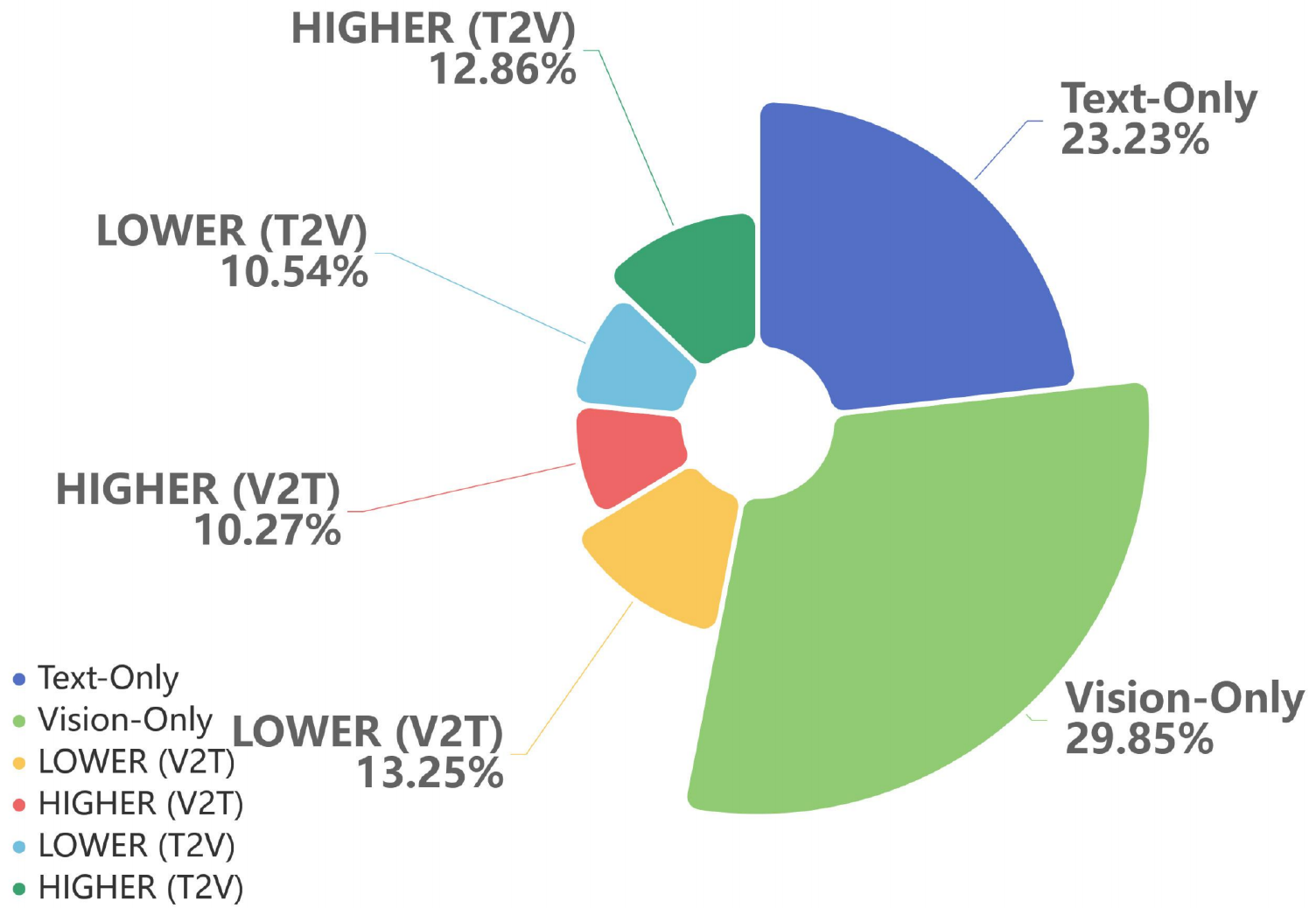}
    \label{subfig:case2}
  }\hfill
  \subfloat[Per/Org/leader of]{
    \includegraphics[width=0.3\linewidth, height=4.5cm, keepaspectratio]
    {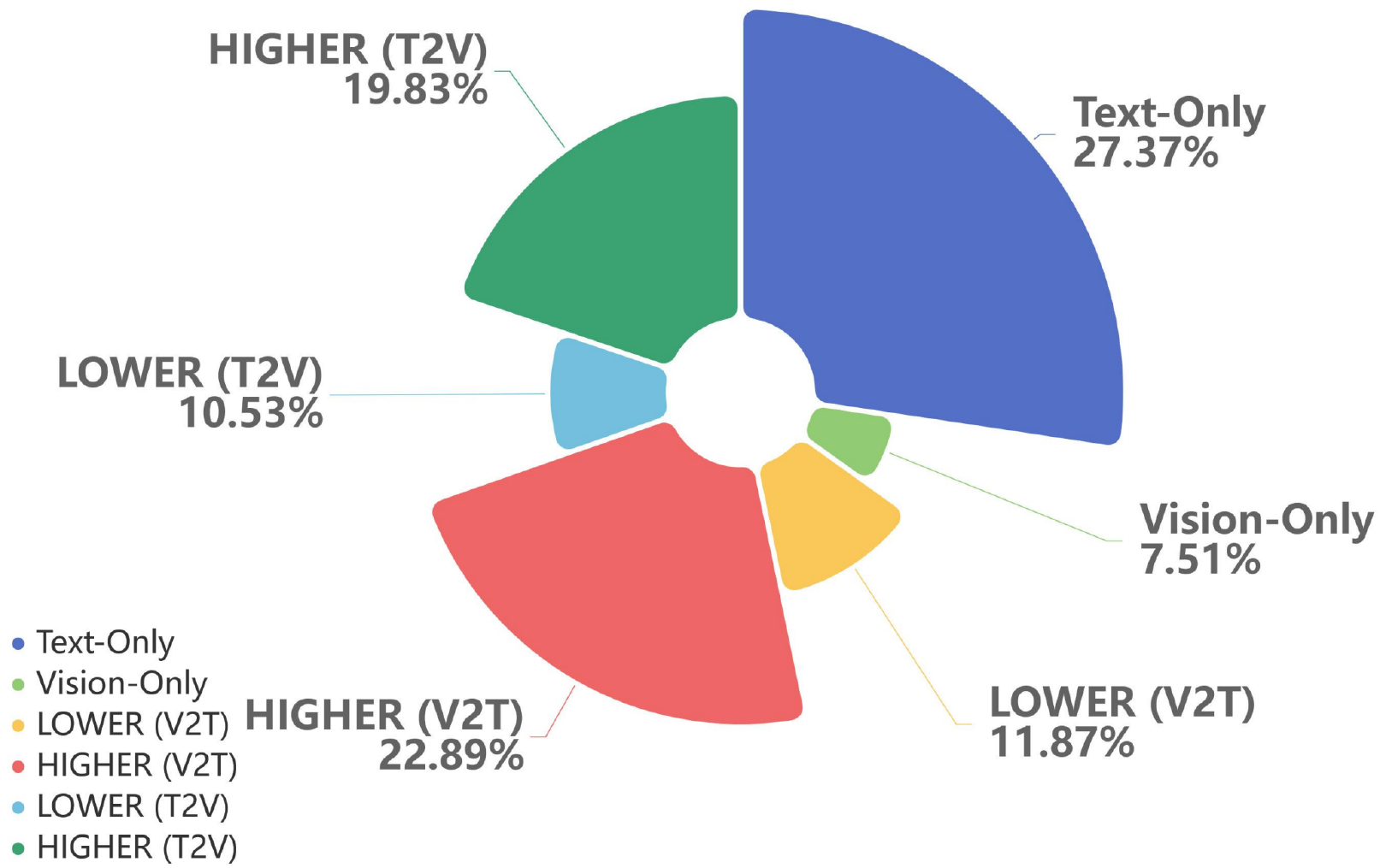}
    \label{subfig:case3}
  }

  \caption{Multimodal Mixture-of-Expert weights visualization results. Lower layers (0-6) capture interaction features containing more environmental context, while higher layers (7-11) specialize in core semantic features of multimodal information.
  }
  
  \label{fig:case_study_2}
\end{figure*}

To verify the effectiveness of each module, we conduct ablation experiments from the perspectives of model structure and components. According to the experimental results listed in Table \ref{table:ablation-results}, we can observe that: 
(1) Replacing Optimal Transport (OT) with Cross Attention (CA) in the MOT module leads to inferior cross-modal alignment, where the source distribution is treated as Key/Value and the target distribution serves as Query. 
While OT enforces global structural matching via Wasserstein distance minimization, CA relies on local similarity metrics that overemphasize dominant features and neglect subtle semantic correlations. The rigid softmax normalization in CA further suppresses weak but meaningful interactions, whereas OT's softmatch strategy inherently maintains topological coherence across modalities.
(2) Removing the MOT module will significantly reduce performance, demonstrating that fusing multilevel features effectively captures fine-grained information crucial for relation extraction. 
(3) Discarding the MMoE module also leads to a performance decline, indicating that different relational triplets emphasize distinct interaction features. The MMoE module can dynamically select the most relevant information from each modality for every triplet.


\begin{table}[ht]
\caption{Feature ablations (P:position, C:caption, D:depth)}
\label{table:ablations}
\centering
\small
\begin{tabular}{ccc|cc}
\hline
 \multicolumn{3}{c|}{\textbf{Feature}}  & \multirow{2}{*}{\textbf{Accuracy}} & \multirow{2}{*}{\textbf{F1-Score}} \\
\cline{0-2}
 P& C& D & & \\
\hline
 &  &  & 74.64  & 63.11 \\
\checkmark & &  & 75.88 & 65.13\\
& \checkmark &  & 76.29 & 66.43\\ 
& & \checkmark &  75.78  & 64.99 \\
\checkmark & \checkmark &  & 77.16 & 67.07 \\
\checkmark & & \checkmark & 76.29 & 65.66 \\
& \checkmark & \checkmark & 76.88 & 66.89 \\
\hline
\checkmark& \checkmark& \checkmark & \textbf{78.47} & \textbf{69.17} \\
\hline
\end{tabular}
\end{table}

As shown in Table \ref{table:ablations}, multimodal features significantly enhance performance on UMRE. Captions are most effective in bridging the modality gap between image and text, while position and depth also contribute, though less prominently. Combining all three achieves the best results, which highlights the importance of leveraging complementary information from multiple modalities for cross-modal reasoning.

\begin{table}[ht]
    \centering
    \small
    \caption{Expert ablations on UMRE (T2V: Text-to-Vision, \\ V2T: Vision-to-Text, T: Text-only, V: Vision-only)}
    \label{tab:moe_ablation}
    \begin{tabular}{cccc|cc}
        \hline
        T2V & V2T & T & V & \textbf{Accuracy} & \textbf{F1-Score} \\
        \hline
        \checkmark & & & & 76.87 & 67.56 \\
        & \checkmark & & & 76.43 & 66.98 \\
        \checkmark & \checkmark & \checkmark & & 77.76 & 68.29 \\
        & \checkmark & \checkmark & \checkmark & 77.14 & 68.13 \\
        \hline
        \checkmark & \checkmark & \checkmark & \checkmark & \textbf{78.47} & \textbf{69.17} \\
        \hline
    \end{tabular}
\end{table}

As shown in Table \ref{tab:moe_ablation}, our approach incorporates multiple specialized experts that collaborate effectively to dynamically select the most relevant interaction features for different relational triplets and relation types.

\begin{figure}[ht]
  \centering
  \Description{A figure comparing model performance under different caption ablation scenarios.} 
  \includegraphics[width=\linewidth]{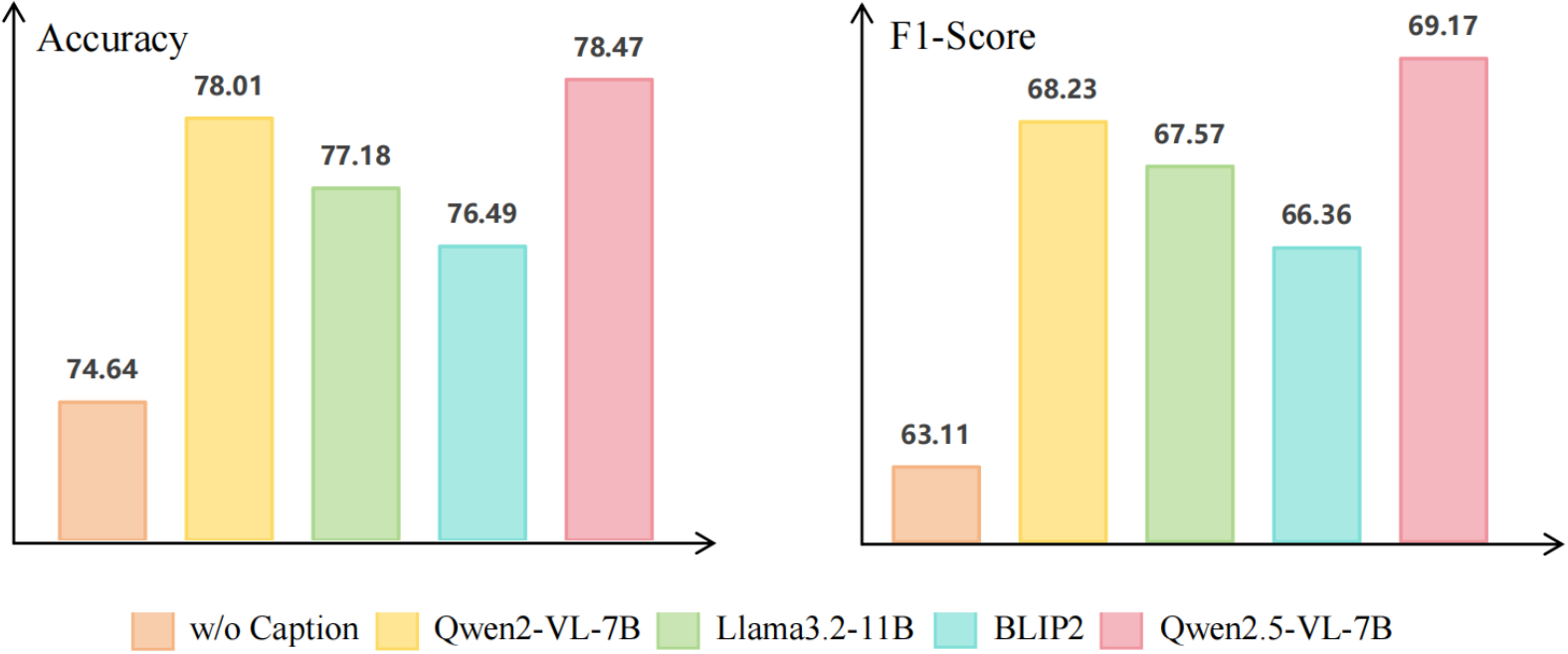}
  \caption{Caption ablations on UMRE}
  \label{fig:caption_ablation}
\end{figure}

Moreover, comparative experiments in image captioning, shown in Fig~\ref{fig:caption_ablation}, reveal that MLLMs with stronger associative capabilities generate superior captions for multimodal relation extraction. This demonstrates that our framework possesses upward compatibility, as its performance will be continuously enhanced with the evolution of MLLMs.

\subsection{Further Analysis}

\begin{table}[ht]
\caption{
    The overall performance on subtask datasets
}
\label{table:sub-results}
\centering
\small 
    \begin{tabular}{c|cc|cc|cc}
    \hline
    \multirow{2}{*}{\textbf{Model}} & \multicolumn{2}{c|}{\textbf{Entity-Entity}} & \multicolumn{2}{c|}{\textbf{Obect-Object}} & \multicolumn{2}{c}{\textbf{Entity-Object}} \\
    \cline{2-7}
    & \textbf{Acc} &  \textbf{F1} & \textbf{Acc} & \textbf{F1} & \textbf{Acc} & \textbf{F1} \\
    \hline
    HVFormer  & 73.23 & {60.86} & 66.87 & 40.22 & 75.87 & 67.08 \\
    MOREformer  & 73.10 & 60.72 & \textbf{69.57} & {43.18}  & 78.84 & 71.67 \\ 
    FocalMRE  & 73.82 & 62.03 & 68.52 & 41.28  & 76.64 & 69.43 \\ 
    \hline
    \text{Ours}  & \textbf{74.63} & \textbf{63.87} & 68.52 & \textbf{45.87}  & \textbf{81.64} & \textbf{74.43} \\
    \hline
    \end{tabular}
\end{table}

Table~\ref{table:sub-results} compares the performance of state-of-the-art models when trained exclusively on individual UMRE subtasks. Our model achieves superior results across all subtasks. The results demonstrate that the MMoE dynamically adapts to select the most suitable features even for triplets of the same type, highlighting its ability to flexibly address diverse feature dependencies in relation extraction. This underscores the module's critical role in modeling cross-modal heterogeneity.

\subsection{Case Study}

\begin{figure}[ht]
  \centering
  \Description{Case study of some examples.} %
  \includegraphics[width=\linewidth, height=0.25\textheight]{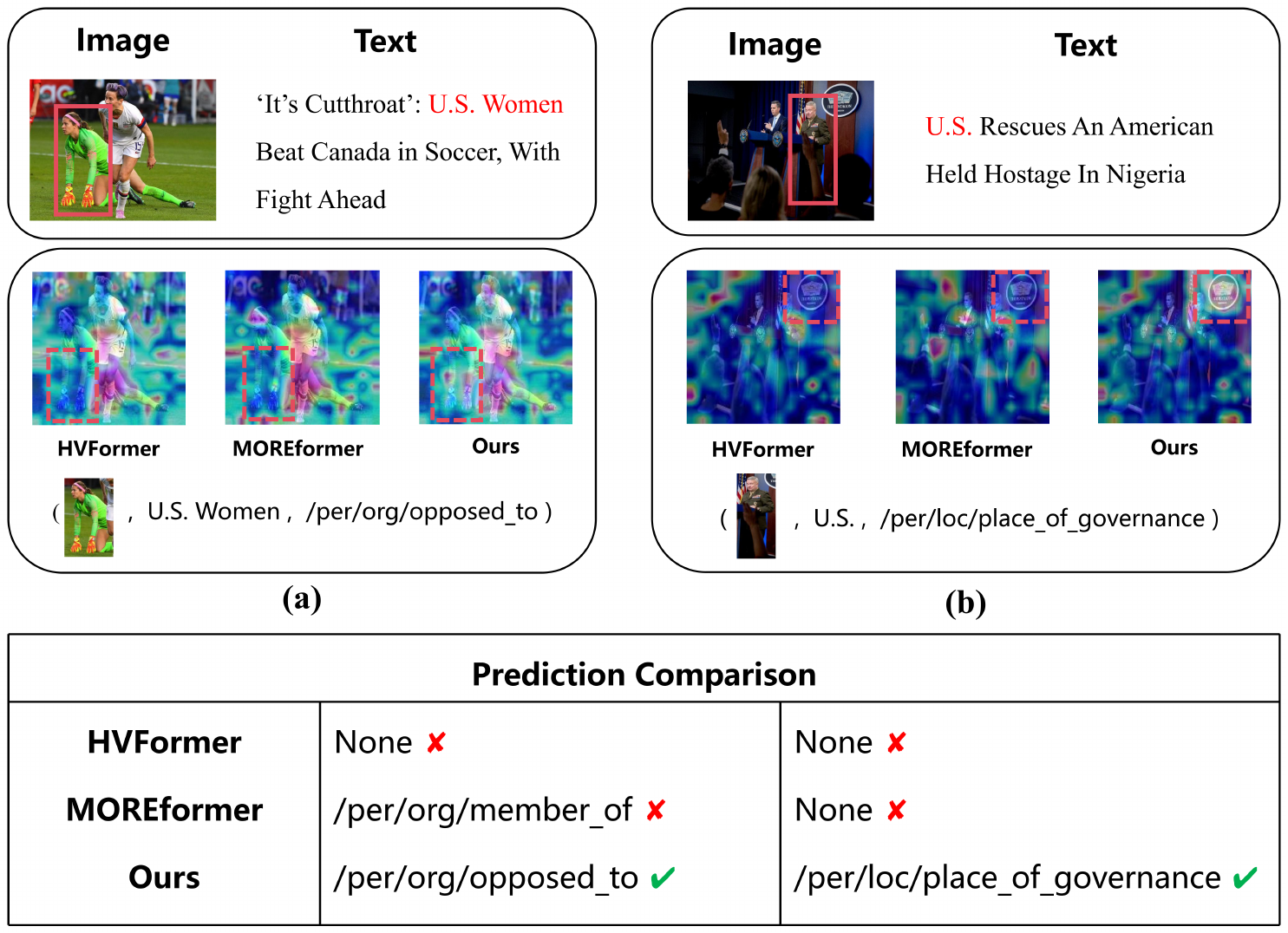}
  \caption{Prediction comparison on two test samples. $\checkmark$ and $\times$ denote correct and incorrect predictions.}
  \label{fig:case_study}
\end{figure}

We conduct a case study comparing REMOTE with other state-of-the-art methods. In Fig. \ref{fig:case_study}(a), our method accurately captures the \textbf{goalkeeper's gloves} and correctly predicts the adversarial relationship between the \textbf{goalkeeper} and the \textbf{U.S. Women's Soccer Team}, while Moreformer and HVFormer fail to do so. In Fig. \ref{fig:case_study}(b), our method identifies the \textbf{Pentagon} and correctly infers the speaker's identity as a \textbf{Government Official}. These results demonstrate that our method effectively retains critical details through multilevel optimal transport feature fusion, leading to more accurate identification of key image regions and their corresponding textual information.

The case study in Fig.~\ref{fig:case_study_2} highlights the effectiveness of our MMoE module in dynamically allocating weights to cross-modal features for different relations.
For spatial-temporal relations such as "held on" (Fig.~\ref{subfig:case1}), the MMoE module prioritizes lower-layer vision-to-text interactions, effectively capturing detailed environmental cues. In contrast, for role-based relations like "president" and "leader of" (Fig.~\ref{subfig:case2}, \ref{subfig:case3}), it emphasizes visual modality features, with "leader of" demonstrating balanced vision-text bidirectional alignment to identify organizational roles. Textual features consistently dominate as the primary semantic anchor across all relations, while visual features provide essential contextual support. These results confirm that the MMoE module effectively balances multiscale evidence integration, enhancing cross-modal reasoning by leveraging the strengths of each modality where they matter most.

\section{Conclusion}

In this paper, we propose a novel framework for Unified Multimodal Relation Extraction, designed to simultaneously extract intra-modal and inter-modal relations between textual entities and visual objects. We also construct a new high-quality dataset for the task. 
Our method leverages multilevel optimal transport feature fusion to capture low-level details overlooked by previous methods and employs multimodal mixture of experts to dynamically select optimal interaction features for different relational triplets.
Extensive experiments on three datasets demonstrate its effectiveness. 
In future work, we will focus on further reducing the modality gap in multimodal fusion.


\begin{acks}
We would like to express our gratitude to the anonymous reviewers for their constructive comments, as well as to all the hardworking and professionally dedicated annotators.
This work was supported by the National Natural Science Foundation of China (No.62406319), and the National Key Laboratory of Science and Technology on Blind Signal Processing (No.23007522).
\end{acks}

\bibliographystyle{ACM-Reference-Format}
\balance
\bibliography{sample-base}










\end{document}